# Improving the estimation of directional area scattering factor (DASF) from canopy reflectance: theoretical basis and validation

Yi Lin [a,1], Siyuan Liu[a,1], Lei Yan[a,b], Kai Yan[c], Yelu Zeng[d], Bin Yang[e,*]

[a] Institute of Remote Sensing and Geographic Information System, School of Earth and Space Sciences, Peking University, Beijing 100871, China. liny@pku.edu.cn (Yi Lin), liusy0527@pku.edu.cn (Siyuan Liu) and lyan@pku.edu.cn (Lei Yan)

[b] Guangxi Key Laboratory of Remote Measuring System, Guiling University of Aerospace Technology, Guilin 541004, China.

[c] School of Land Science and Techniques, China University of Geosciences, Beijing 100083, China. kaiyan@cugb.edu.cn (Kai Yan)

[d] College of Land Science and Technology, China Agricultural University, Beijing, 100091, China. zengyelu@163.com (Yelu Zeng)

[e] College of Electrical and Information Engineering, Hunan University, Changsha 410082, China. binyang@hnu.edu.cn (Bin Yang)

[1] These authors contributed equally.

**Author for correspondence**
Bin Yang (binyang@hnu.edu.cn)
Submitted for publication in Remote Sensing of Environment

**Abstract:**

Directional area scattering factor (DASF) is a critical canopy structural parameter for vegetation monitoring. It provides an efficient tool for decoupling of canopy structure and leaf optics from canopy reflectance. Current standard approach to estimate DASF from canopy bidirectional reflectance factor (BRF) is based on the assumption that in the weakly absorbing 710 to 790 nm spectral interval, leaf scattering does not change much with the concentration of dry matter and thus its variation can be neglected. This results in biased estimates of DASF and consequently leads to uncertainty in DASF-related applications. This study proposes a new approach to account for variations in concentrations of this biochemical constituent, which additionally uses the canopy BRF at 2260 nm. *In silico* analysis of the proposed approach suggests significant increase in accuracy over the standard technique by a relative root mean square error (rRMSE) of 49% and 34% for one- and three dimensional scenes, respectively. When compared with indoor multi-angular hyperspectral measurements reported in literature, the mean absolute error has reduced by 68% for needle leaf and 20% for broadleaf canopies. Thus, the proposed DASF estimation approach outperforms the current one and can be used more reliably in DASF-related applications, such as vegetation monitoring of functional traits, dynamics, and radiation budget.

# 1. Introduction

The vegetation shortwave radiation budget plays a critical role in understanding the near-surface climate as it describes the partitioning of incident radiation into canopy absorbed and scattered portions (Adams et al. 2018; Huang et al. 2007; Ollinger et al. 2008; Stenberg et al. 2016). The interaction between incident radiation and vegetation canopy is determined by canopy structure, leaf optical properties and reflectivity of the ground. Despite that the possibility of a photon being absorbed varies with wavelength (which is determined by the leaf optical properties), the interaction probability is spectrally invariant and is solely determined by canopy structure (Huang et al. 2007). This property triggers the spectral invariants theory, i.e., canopy absorption and scattering are explicit functions of wavelength-independent variables, including, among others, canopy interceptance and recollision probability, and wavelength-dependent leaf albedo (Huang et al. 2007; Lewis and Disney 2007; Smolander and Stenberg 2005). The spectral invariants theory simply and effectively describes photon-canopy interactions. In this way, spectral invariants theory differs from many other canopy radiative models where more complicated radiation transfer processes are incorporated (Stenberg et al. 2016).

Spectral invariants have been widely used to accurately reconstruct the radiation regime at both canopy (Huang et al. 2007; Rautiainen and Stenberg 2005; Zeng et al. 2018) and leaf (Wu et al. 2021) scales, laying the foundation for the estimation of these variables from remotely sensed data (Knyazikhin et al. 2011; Lukes et al. 2011; Schull et al. 2007). Studies have shown success in spectral-invariant-based estimations of canopy structural parameters such as leaf area index (LAI)(Rautiainen et al. 2009; Yan et al. 2016; Yan et al. 2021a; Yan et al.

2021b), clumping index (CI)(Smolander and Stenberg 2003) and canopy height(Schull et al. 2007). Spectral invariants also provide a useful tool to decouple information regarding canopy structure and leaf optical properties conveyed in the canopy bidirectional reflectance factor (BRF), which consequently can be used for the accurate estimation of foliar biochemical concentrations (Knyazikhin et al. 2013; Ustin 2013). These applications were largely implemented by the conception of the directional area scattering factor (DASF)(Knyazikhin et al. 2013). DASF is defined as the canopy BRF if the foliage does not absorb radiation and it is correlated with the near-infrared (NIR) reflectance of vegetation (Knyazikhin et al. 2013; Zeng et al. 2018). It is physically interpreted as an estimate of the ratio between the leaf area that forms the canopy boundary as seen along a given direction and the total one-sided leaf area (Knyazikhin et al. 2013). DASF is an important structural parameter for agriculture and vegetation monitoring. On the one hand, DASF is closely related to various vegetation structural parameters, such as the broadleaf fraction of forests (Vanhatalo et al. 2014) and CI (Stenberg and Manninen 2015). Provided the close relationship between DASF and LAI, DASF has also been utilized for retrieving LAI and its sunlit fraction from DSCOVR EPIC and Sentinel-2 MSI data (Brown et al. 2021; Yang et al. 2017). On the other hand, DASF can be used to correct for structural effects from canopy reflectance and give a more reliable estimation of canopy scattering coefficients, which are closely related to foliar biochemistry. Therefore, DASF helps to accurately retrieve leaf biochemical concentrations (Latorre-Carmona et al. 2014; Liu et al. 2020; Wang et al. 2017). Besides these applications, research utilizing DASF for describing vegetation dynamics, productivity and the radiation budget in the global radiation balance or climate models are undergoing or expected to arise in the near future (Kohler et al. 2018; Stenberg et al. 2016). Therefore, the accurate estimation of DASF from remotely sensed data is required for SI-based studies on a wide range of vegetation functional traits and other environmental parameters.

For vegetation canopies with a dark background or for sufficiently dense vegetation where the impact of the canopy background is negligible, DASF can be directly obtained given canopy BRF and average leaf albedo (Adams et al. 2018; Vanhatalo et al. 2014). In reality, leaf albedo is usually not known in remote sensing. To tackle this problem, Knyazikhin et al. (2013) proposed a practical DASF estimation method with the help of a reference leaf albedo (i.e., the *sDASF* algorithm). It is efficient and quite easy to use. However, this algorithm assumes that leaf albedo does not change much with the concentration of dry matter in 710 to 790 nm spectral interval (Knyazikhin et al. 2013; Yang et al. 2016), and consequently neglects its variation when estimating DASF. This introduces uncertainty in the estimation of DASF (see Appendix A). Although Adams et al. (2018) concluded that DASF can be accurately extracted using the reference leaf albedo with various viewing geometries, 3-D arrangements of leaves and different soil conditions, they kept the above assumption when simulating leaf albedo. Note that DASF has been widely used in many related works, such as generation of look-up-table for DSCOVR LAI products (Yang et al. 2017). Comprehensive assessments of the influence of the assumption on DASF estimation and the consequent uncertainty in related applications has not yet been conducted.

To address these concerns, the objectives of this study are: 1) to build a model that compensates for the bias of DASF estimation by accounting for the variation of green leaf albedo with dry matters, and to obtain an improved DASF estimation algorithm that incorporates the correction model (i.e., the *iDASF* algorithm); 2) to assess the performance of the *iDASF* algorithm over homogeneous and heterogeneous vegetation and to analyze the model sensitivity to various canopy structural parameters.

This paper is organized as follows. Section 2 presents the theoretical basis of the bias caused by the *sDASF* algorithm. An improved DASF estimation algorithm incorporated with a correction model for the bias is proposed in this section. Validation of the improved approach over simulated one-dimensional (1-D) and three-dimensional (3-D) vegetated surfaces are described in Section 3. An analysis of the *iDASF* algorithm to canopy structural properties, e.g. viewing geometry, LAI and leaf inclination distribution function (LIDF), is also presented. In Section 4, we present further validation of the *iDASF* algorithm using a dataset from observations. Conclusions are given in Section 5.

## 2. Theoretical basis of compensation for DASF estimation

### 2.1 Canopy BRF

BRF of a vegetation canopy, which is bounded below by a black surface, can be parameterized by a set of spectrally invariant variables, namely, canopy interceptance, $i_0$, recollision probability, $p$, and directional escape density, $\rho(\Omega)$, and an optical variable, i.e., leaf albedo, $\omega_\lambda$, as (Huang et al. 2007; Knyazikhin et al. 2011):

$$BRF_\lambda(\Omega) = \frac{\rho(\Omega)i_0}{1-\omega_\lambda p}\omega_\lambda = DASF \cdot W_\lambda, \tag{1}$$

where:

$$DASF = \frac{\rho(\Omega)i_0}{1-p}. \tag{2}$$

$$W_\lambda = \frac{1-p}{1-\omega_\lambda p}\omega_\lambda. \tag{3}$$

Here, $i_0$ is the fraction of incident radiation that is intercepted by the vegetation canopy, $p$ is the probability that a photon scattered by a leaf in the canopy will interact with another leaf in the canopy again (Huang et al. 2007; Smolander and Stenberg 2005), and $\rho(\Omega)$ is the probability that a photon scattered by a leaf in the canopy will escape the canopy in the direction $\Omega$ (Stenberg 2007). The three spectral invariants can be combined into DASF (Equation (2)). This variable is totally determined by the canopy structure and is insensitive to leaf biochemical concentrations. The canopy scattering coefficient, $W_\lambda$, is determined by the recollision probability and the leaf albedo. Compared to BRF, $W_\lambda$ is more closely related to leaf biochemical concentrations (Latorre-Carmona et al. 2014).

**2.2 Bias of the standard DASF algorithm**

Theoretical and empirical analyses suggest that DASF can be directly retrieved from the canopy BRF in the 710 to 790 nm spectral region, without involving any canopy radiative transfer models or prior knowledge regarding leaf scattering properties. For a vegetation canopy bounded below by a non-reflecting background, the standard approach for retrieval of DASF (i.e.. the *sDASF* algorithm) is as follows (Knyazikhin et al. 2013):

(1) Calculate the reference leaf albedo, $\omega_r$, using the PROSPECT model(Jacquemoud and Baret 1990) with leaf chlorophyll concentration ($C_{ab}$) 16 $\mu g/cm^2$, equivalent water thickness (EWT) 0.005 $cm^{-1}$ and leaf dry mass per unit area (LMA) 0.002 $g/cm^2$;

(2) Plot the ratio of BRF/$\omega_r$ versus BRF in the 710-790 nm spectral region. Let $k$ and $b$ be the slope and intercept of the BRF/$\omega_r$ versus BRF relationship;

(3) Estimate the canopy DASF by $b/(1-k)$.

The above *sDASF* algorithm is effective and quite easy to use. However, it was proposed based on the assumption that variation of leaf scattering albedo with concentration of leaf dry matter is neglected in 710 to 790 nm spectral interval, which cannot be always satisfied and leads to bias estimates of DASF (see Appendix A). This problem will be fixed in the following parts.

In fact, in 710 to 790 nm spectral region, chlorophyll and dry matter are the two dominant absorbing foliar constituents, while water absorption can be neglected (Knyazikhin et al. 2013). Let $t_c$ and $t_m$ be the ratios between the $C_{ab}$ and LMA of a green leaf and the reference $C_{ab}$ and LMA, respectively. Following the *sDASF* algorithm, one obtains the biased DASF, named as $DASF'$, as (see Appendix A):

$$DASF' = \frac{D\rho(\Omega)i_0}{D(1-p) + DC} = \frac{\rho(\Omega)i_0}{1 - p + C}, \tag{4}$$

where:

$$C = \frac{1-A}{A(1-p_{leaf})}, \tag{5}$$

$$A = \exp(-(t_d - t_m)C_m k_m), \tag{6}$$

$$D = A[1 - q(t_c)]. \tag{7}$$

Here, $C_m$ and $k_m$ are the reference concentration of LMA and its corresponding modified specific absorption coefficient, which is flat in the 710 to 790 nm spectral region. $p_{leaf}$ is the recollision probability between the within-leaf fundamental terms. (Lewis and Disney 2007). $q(t_c)$ equals to $(t_c - 1)/t_c$, see Appendix A. The comparison between Equation (4) and Equation (2) indicates that *sDASF* gives a biased estimation of DASF (i.e., $DASF'$), and $DC$ is the bias factor.

## 2.3 Bias estimation using radiative transfer model

To correct the bias in the *sDASF* algorithm, the real DASF ($DASF_0$) can be calculated as the ratio between the intercept and (1-slope-*DC*), i.e.:

$$DASF_0 = \frac{b}{1-k-DC} = \frac{D\rho(\Omega)i_0}{D(1-p)+DC-DC} = \frac{\rho(\Omega)i_0}{1-p}. \tag{8}$$

The parameter *DC* is the key for the correction of $DASF'$. Following Equation (5) to (7), *DC* can be expressed as:

$$DC = \frac{1-\exp[(t_c-t_m)C_m k_m]}{t_c(1-p_{leaf})}. \tag{9}$$

*DC* is determined by the green leaf $C_{ab}$, LMA, and $p_{leaf}$. Therefore, the key to the improved DASF algorithm is to find an effective way for compensating *DC* from remote sensing data.

Canopy BRF can be used for *DC* modeling because $C_{ab}$, LMA and within-leaf recollision probability are conveyed in the vegetation reflection spectrum. Previous studies suggested that the red-edge canopy BRF is well correlated with $C_{ab}$ (Clevers and Kooistra 2012; Verrelst et al. 2018). In the red-edge spectral region, leaf albedo at wavelengths below 720 nm is mainly determined by the absorption of chlorophyll (Fig. 1). Consequently, $C_{ab}$ can be well reflected by the red-edge canopy BRF below 720 nm (e.g., 710 nm BRF)(le Maire et al. 2008; Verrelst et al. 2016). LMA strongly absorbs radiation from 1400 nm to 2400 nm (Feret et al. 2021), which overlaps with the absorbing spectral region of water. However, 2260 nm corresponds to a better compromise between the relatively low absorption of water and the high absorption of dry matter (Fig. 1), and is well correlated with LMA at the canopy level (le Maire et al. 2008). In this study, BRF at 710 nm (BRF$_{710}$) and 2260 nm (BRF$_{2260}$) were, thus, respectively used as indicators of $C_{ab}$ and LMA and, accordingly, for modeling $DC$.

To explore the relationship among $DC$, BRF$_{710}$, and BRF$_{2260}$, we designed an experiment based on numerous green leaves with known biochemical concentrations and the corresponding canopy BRF. The former can be obtained by generating a synthetic dataset, and the latter can be simulated using a canopy RTM, e.g., PROSAIL (Jacquemoud et al. 2009). A synthetic dataset consisting of 2000 combinations of four foliar biochemical contents, i.e., $C_{ab}$, carotenoid concentration (Car), EWT, and LMA, was generated to simulate 2000 green leaves. The synthetic data were generated using a multivariable normal distribution based on the maximum and minimum values, mean values and standard deviations of the concentrations of the four biochemical contents as well as their correlations (Tab. 1), which were calculated from numerous leaf samples (Feret et al. 2011). Such simulation of green leaves has been shown to be a good sampling strategy as it includes the correlation between leaf constituents and thus avoids generating unrealistic combinations (Feret et al. 2011). It is notable that in these simulated leaves, leaves with $C_{ab}$ less than 10 µg/cm² were excluded as only green leaves were considered in this study. Finally, 1932 combinations were selected as a result.

The PROSPECT-D model (Féret et al. 2017) was further coupled with the SAIL model (Verhoef 1984) (PROSAIL-D) to calculate canopy BRF based on the synthetic data. Input parameters required in the PROSAIL-D are listed in Tab. 2. In this study, leaf anthocyanins and brown pigments concentrations were set to zero, and only the black soil case (with non-reflecting soil bounded below) was considered.

Tab. 1 Correlation matrix for the four types of leaf biochemical constituents, i.e. Chlorophyll a and b ($C_{ab}$), Carotenoid (Car), leaf mass per area (LMA) and equivalent water thickness (EWT) in the synthetic data. Numbers in the parentheses are the correlation coefficients presented in Feret et al. (2011).

|  | $C_{ab}$ | Car | LMA | EWT |
|---|---|---|---|---|
| $C_{ab}$ | 1 | | | |
| Car | 0.85 (0.86) | 1 | | |
| LMA | 0.19 (0.19) | 0.42 (0.43) | 1 | |
| EWT | 0.19 (0.19) | 0.26 (0.27) | 0.63 (0.63) | 1 |

Tab. 2 The input parameters of PROSAIL-D for the simulation of canopy BRF.

| Parameter | Abbrev. | Unit | Value |
|---|---|---|---|
| Leaf structure parameter | N | - | 1.5 |
| Leaf chlorophyll a and b concentration | $C_{ab}$ | μg/cm² | * |
| Leaf carotenoid concentration | Car | μg/cm² | * |
| Leaf dry matter per area | Cm | g/cm² | * |
| Leaf equivalent water thickness | EWT | $cm^{-1}$ | * |
| Leaf area index | LAI | m²/m² | 5 |
| Hot spot effect | - | - | 0.01 |
| Solar zenith angle | SZA | degree | 30 |
| View zenith angle | VZA | degree | 0 |
| Relative azimuth angle between solar and view directions | RAA | degree | 0 |
| Leaf inclination distribution function | LIDF | - | a=0, b=0 (uniform) |

\* values are from the synthetic dataset.

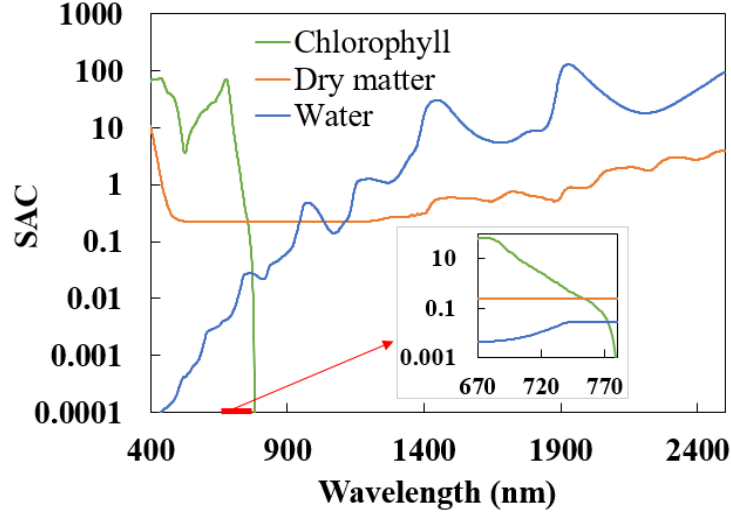

Fig. 1 Specific absorption coefficient (SAC) of chlorophyll (in cm$^2$/mg), dry matter (in cm$^2$/dg) and water (in cm$^{-1}$) from the SAC library of PROSPECT-D. In these units typical concentrations of these three constituents are of comparable magnitudes (Knyazikhin et al. 2013).

Let $DC_0$ denote the bias produced by *sDASF*. For a given canopy with known average leaf albedo $\omega_\lambda$, $\rho(\Omega)i_0$ and $p$ can be directly obtained as the intercept $b_0$ and the slope $k_0$ of the linear relationship between BRF/$\omega_\lambda$ and BRF (Equation (1)) (Knyazikhin et al. 2011; Lukes et al. 2011; Schull et al. 2011). DASF can be thus obtained by $DASF_0 = b_0/(1 - k_0)$ (Adams et al. 2018). This allows us to obtain $DASF_0$ and consequently to calculate $DC_0$ using the PROSAIL-D model with the synthetic dataset. $DC_0$ can be obtained following Equation (8) as:

$$DC_0 = 1 - k - \frac{b}{DASF_0}, \qquad (10)$$

where $k$ and $b$ are the slope and intercept of BRF/$\omega_r$ versus the BRF linear relationship, respectively.

In total, 1932 $DC_0$-BRF$_{710}$-BRF$_{2260}$ combinations could be generated. As shown in Fig. 2, $DC_0$ ranges from -0.05 to 0.2. Positive $DC_0$ suggested an underestimation of DASF using *sDASF*, and vice versa. According to Fig. 2 (a), larger $BRF_{710}$ and lower $BRF_{2260}$ yield higher $DC_0$, indicating that the *sDASF* algorithm is more likely to underestimate DASF over vegetation with lower $C_{ab}$ and higher LMA given that $C_{ab}$ and LMA are negatively related to $BRF_{710}$ and $BRF_{2260}$, respectively. Although it is hard to directly build a 3-D model between $DC_0$ and 710 nm and 2260 nm BRF, an explicit exponential relationship can be obtained after rotating the 3-D view (Fig. 2(a)) to a specific 2-D view (Fig. 2(b)), where the *y* axis remains DC whereas the *x* axis transforms to a linear combination of $BRF_{710}$ and $BRF_{2260}$. Optimal coefficients for the linear combination could be achieved by minimizing the root mean square error (RMSE). Fig. 2(b) shows the view for the optimal regression. After substitution of the linear expression of the *x*-axis with the exponential formula shown in Fig. 2(b), *DC* can be modeled as:

$$DC = \exp(9.3894 BRF_{710} - 15.1453 BRF_{2260} - 3.5058) - 0.0227. \qquad (11)$$

The modeled *DC* correlates well with $DC_0$ (Fig. 3). According to Equation (11), one can easily derive *DC* if the canopy BRF at 710 nm and 2260 nm are known.

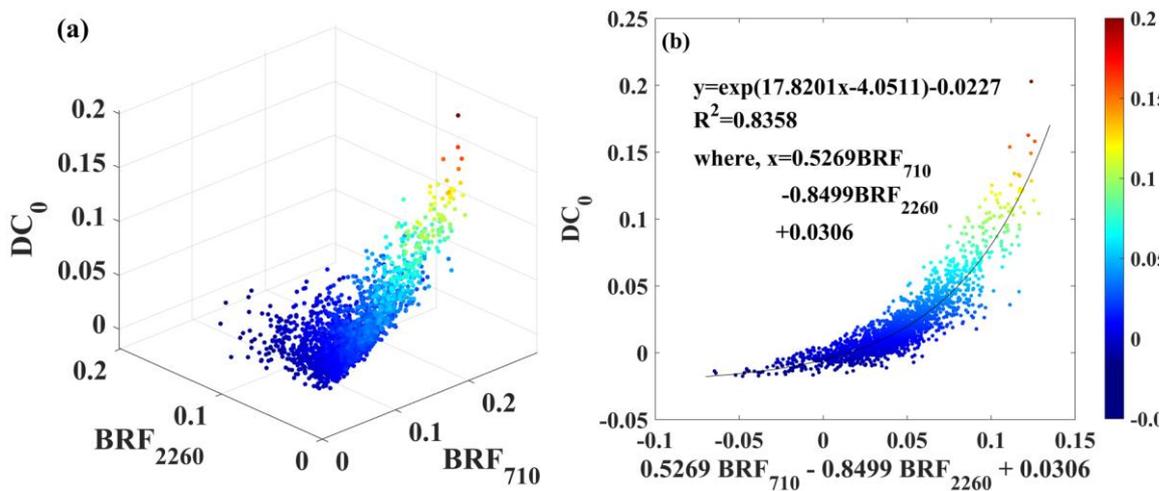

Fig. 2 Variation of $DC_0$ with BRF at 710nm ($BRF_{710}$) and at 2260 nm ($BRF_{2260}$) in 3-D view (a) and its rotated 2-D view (b) of optimal regression. The solid line in panel (b) represents the exponential

regression line of the scatters.

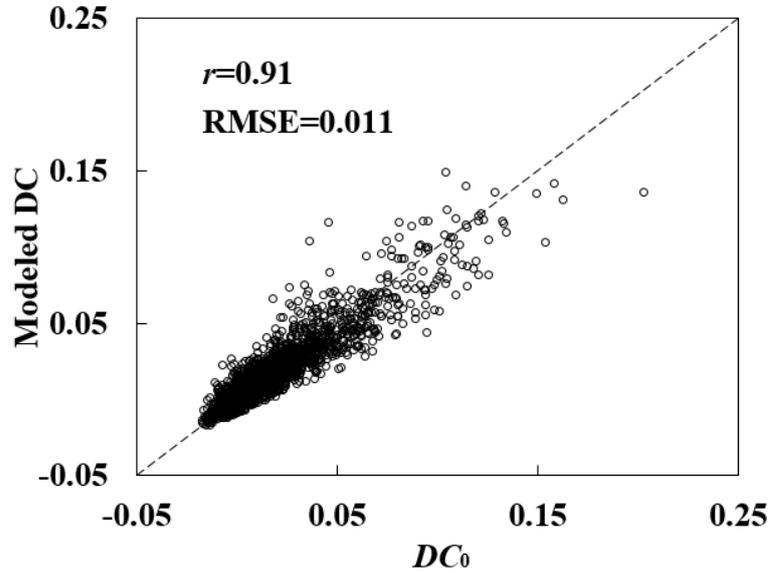

Fig. 3 Correlation between $DC_0$ and modeled $DC$ from Equation (11).

**2.4 Improved estimation of DASF**

Once $DC$ is estimated via Equation. (11), the corrected DASF can be easily calculated as:

$$DASF = \frac{b}{1 - k - DC}, \tag{12}$$

and the improved algorithm for DASF estimation (*iDASF*) can be described as follows:

(1) Calculate the reference leaf albedo $\omega_r$ using the PROSPECT model with $C_{ab}$ 16 $\mu g/cm^2$, EWT 0.005 $cm^{-1}$ and LMA 0.002 $g/cm^2$;

(2) Plot the ratio BRF/$\omega_r$ versus BRF in the 710-790 nm spectral region, with *k* and *b* being the slope and intercept of the BRF/$\omega_r$ versus BRF relationship;

(3) Calculate $DC$ using the canopy BRF at 710 nm and 2260 nm following Equation (11);

(4) Estimate the corrected DASF as $b/(1 - k - DC)$.

Compared to *sDASF* described in Section 2.2, *iDASF* involves one more step for estimating *DC*. Note that *iDASF* incorporated with the *DC* modeled from Equation (11) is exactly based on the specified reference leaf albedo with the biochemical configurations listed in step (1). The *iDASF* (Equation (12)) is thus capable of accurately estimating DASF by accounting for the variation of dry matter on leaf albedo.

## 3. Validation of *iDASF* using radiative transfer models

By considering the variation of both $C_{ab}$ and LMA for different green leaves, the *iDASF* algorithm was proposed for more accurate estimation of DASF. The model was built based on a homogeneous vegetation layer with fixed structural parameters and viewing direction (Tab. 2) using a 1-D RTM (i.e., PROSAIL-D). For wider applicability, validation and sensitivity analyses of *iDASF* along various structural and observing parameters, as well as the model performance on a 3-D heterogeneous vegetated scene were investigated in this section. Studies on 1-D homogenous and 3-D heterogeneous scenes were implemented by PROSAIL-D (Section 3.1) and the Discrete and Anisotropic Radiative Transfer (DART) model(GastelluEtchegorry et al. 1996) (Section 3.2), respectively.

**3.1 Validation over homogeneous vegetation using PROSAIL-D**

LAI and LIDF are the two configurable canopy structural parameters in PROSAIL that have a significant impact on both canopy BRF and DASF. For sensitivity analyses, we sampled LAI from 1 to 7 with a step of one to represent sparse to dense canopies and used six types of LIDF (planophile, erectophile, plagiophile, extremophile, spherical and uniform) described by six combinations of two parameters, *a* and *b* (Verhoef 1997). As for the viewing directions, we considered VZAs from 0° to 60° with a step of 10° in the principal plane (RAA=180°) to represent nadir to oblique observations. When one of the three parameters (LAI, LIDF and VZA) was studied, the other two kept the default value listed in Tab. 2, that is LAI=5, LIDF=uniform, and VZA=0°. Other input parameters of PROSAIL-D coincided with those listed in Tab. 2.

Fig. 4 illustrates the comparison of *DASF*$_0$, DASF retrieved from *sDASF* and from *iDASF* over various LAIs, LIDFs and VZAs, respectively. DASF responds to different structural properties. *DASF*$_0$ rises from 0.2 to 0.6 when LAI increases from 1 to 7; an erectophile LIDF leads to lower *DASF*$_0$ compared with other LIDFs. This can be explained with the observation that the lower LAI and nadir view of an erectophile canopy result in less visible leaf area at the canopy boundary, and thus in lower DASF. For a canopy with uniformly distributed leaves, there is a near-constant fraction of leaves along various viewing angles. In this case, DASF shows little dependence on VZA (Fig. 4(c)). Fig. 4 also illustrates that *DASF*$_0$ are distributed very closely to their mean values, showing little dependence on foliar biochemical properties.

Fig. 4 and Tab. 3 show that *iDASF* performs much better than *sDASF* on a continuous subset of the synthetic data, while *sDASF* underestimates DASF in most cases. We evaluated model performance with the relative root mean square error (rRMSE), which describes the relative error in its root mean square form. Fig 4 shows that for *sDASF*, the deviation from *DASF*$_0$ increases with LAI, with rRMSE rising from 6.76% (LAI=1) to 14.83% (LAI=7). However, the mean estimates from *iDASF* are much closer to *DASF*$_0$, with the rRMSE varying from 3.98% to 7.63% along various LAI, reduced by 50% on average compared to *sDASF*. The same results can be found for the sensitivity analyses of LIDF and VZA, with the rRMSE reduced by 46% and 50% on average, respectively, when *iDASF* was adopted, as presented in Tab. 3.

It is also notable that DASF equals the canopy BRF if $\omega_\lambda$ equals one. This can be validated with an RTM but is impractical for measurements. For a structurally fixed canopy with non-absorbing leaves, the leaf surface optical property is the only factor influencing the canopy BRF. Accounting for this, the refractive index of the wax-air interface on the leaf surface was fixed to 1.5 (Vanderbilt and Grant 1985; Waquet et al. 2009) to obtain a constant non-absorbing canopy BRF. Fig. 4 shows that, in spite of the varying canopy structural properties, DASF retrieved from *iDASF* is consistently closer to the real DASF, i.e., to the non-absorbing canopy BRF.

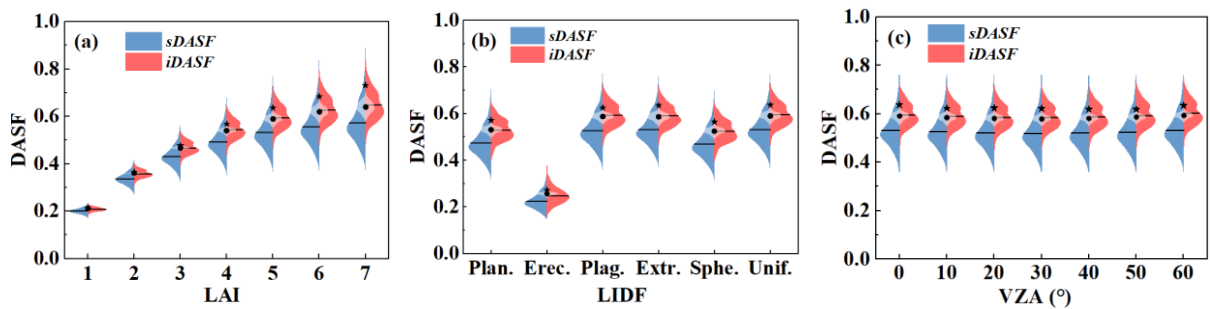

Fig. 4 Split violin distribution of DASF retrieved from the *sDASF* and *iDASF* algorithms for various LAIs (a), LIDFs (b) and VZAs (c). The horizontal lines of the left and right violins denote the mean values. The transparent violin distributions between two split violins represent $DASF_0$, which was calculated in Section 2.3. The black dots represent the mean values of $DASF_0$. The black stars denote non-absorbing canopy BRF.

Tab. 3 rRMSE of DASF retrieved from the *sDASF* and *iDASF* algorithms.

| rRMSE (%) | LAI | | | | | | |
|---|---|---|---|---|---|---|---|
| | 1 | 2 | 3 | 4 | 5 | 6 | 7 |
| *sDASF* | 6.76 | 9.01 | 11.03 | 12.61 | 13.71 | 14.41 | 14.83 |
| *iDASF* | 3.98 | 4.31 | 5.14 | 6.06 | 6.80 | 7.31 | 7.63 |
| | LIDF | | | | | | |
| | Plan. | Erec. | Plag. | Extr. | Sphe. | Unif. | |
| *sDASF* | 14.71 | 17.12 | 14.20 | 13.09 | 14.77 | 13.71 | |
| *iDASF* | 7.32 | 12.61 | 7.08 | 6.42 | 7.39 | 6.80 | |
| | VZA (°) | | | | | | |
| | 0 | 10 | 20 | 30 | 40 | 50 | 60 |
| *sDASF* | 13.71 | 13.99 | 14.25 | 14.48 | 14.64 | 14.65 | 14.41 |
| *iDASF* | 6.80 | 6.93 | 7.08 | 7.24 | 7.39 | 7.49 | 7.53 |

The applicability of *iDASF* on canopies with a variety of structures demonstrates the insensitivity of *DC* to vegetation structural parameters. As illustrated in Fig. 5, the values of modeled *DC* (Equation (11)) follow those of $DC_0$ along different canopy structural parameters configurations. The sensitivity of the *DC* model can be attributed to the sensitivity of $BRF_{710}$ and $BRF_{2260}$ to the variation of the structural parameters. Nevertheless, such sensitivity does not yield a great impact on the performance of *iDASF* along various structure configurations (Fig. 4 and Tab. 3).

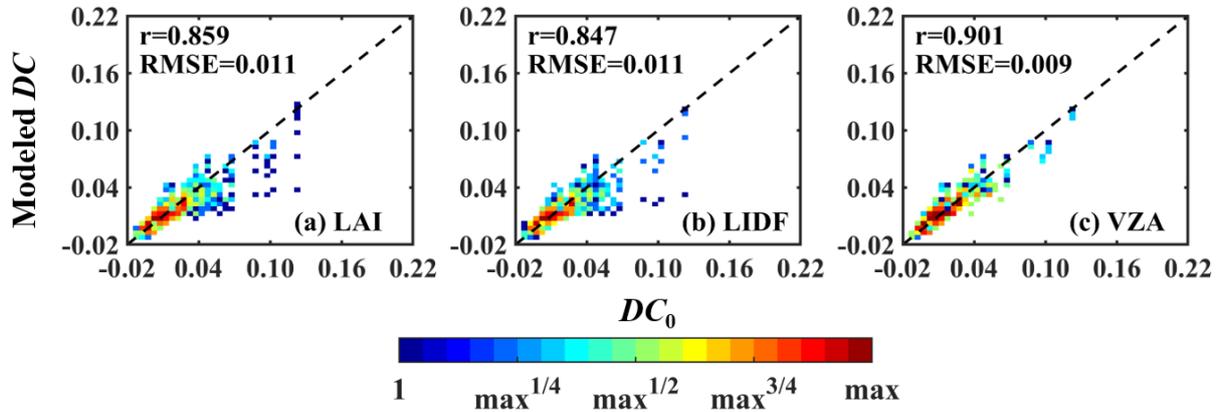

Fig. 5 Scatter plots of $DC_0$ and modeled *DC* for varying LAI (a), LIDF (b) and VZA (c) over 1-D

homogeneous vegetation. Each panel illustrates results from all the values of the corresponding structure parameter that were listed in Tab. 3. The values in each colored bin represent the number of points on a logarithm scale, so that bins with warmer colors indicate many more points than bins with cooler colors.

## 3.2 Validation over heterogeneous vegetation using DART

In this section, the *iDASF* algorithm is further validated over heterogeneous vegetation. The DART model was used in this section for simulating BRF of a heterogeneous vegetated scene. The DART-Lux mode was applied for a more accurate and less time-consuming flux tracking process, as this mode adapts a bidirectional path tracing algorithm that solves the light transport equation with Monte-Carlo integration techniques (Wang and Gastellu-Etchegorry 2021). The version of the model was 5.7.9. Some of the specified model parameters in the DART-Lux mode are listed in Tab. 4.

Tab. 4 Specified parameters in DART-Lux mode.

| Category | Parameter | Value |
| --- | --- | --- |
| Atmosphere | | No atmosphere radiative transfer |
| Flux tracking | Propagation threshold | 1.0E-5 |
| | Smaller mesh size of BOA irradiance sources (m) | 0.125 |
| | Number of iterations | 20 |
| | Target pixel size (meters) | 1 |
| LuxCoreRender | Maximum rendering time per image (seconds) | 0 |
| | Target ray mean density per pixel | 50 |
| | Periodic save time | 20 |
| | Repetitive scene | 1 |

A 10 m × 10 m vegetated sub-scene (Fig. 6) was selected from the standard heterogeneous vegetated scene *HET20* used in the fourth and fifth phases of RAdiation transfer Model Intercomparison (RAMI-IV(Widlowski et al. 2013) and –V). *HET20* describes a 101 m×101 m vegetated scene including 5093 randomly distributed vegetation spheres. Each sphere has a radius of 0.5 m and an LAI of 5, consisting of 49999 discs (scatterers) of negligible thickness with a radius of 0.005 m. The discs are uniformly distributed in the sphere. More details regarding RAMI and *HET20* can be found at https://rami-benchmark.jrc.ec.europa.eu/. The selected 10 m×10 m area is composed of 51 vegetation spheres. To simulate dense vegetation, the size of each sphere was doubled compared with that of *HET20*, i.e., the radius of each sphere and each disc were set to 1 m and 0.01 m, respectively (Fig. 6). The number of scatterers remained the same so that the LAI of each sphere was the same as that of *HET20*. The optical properties of each disc were modeled by PROSPECT-D and followed the bi-Lambertian scattering pattern. The scene was bounded below by black soil. For the purpose of sensitivity analyses on various canopy structural parameters, spheres with different LAI and LIDF were generated using the "Creation of 3D objects" tool of the DART software.

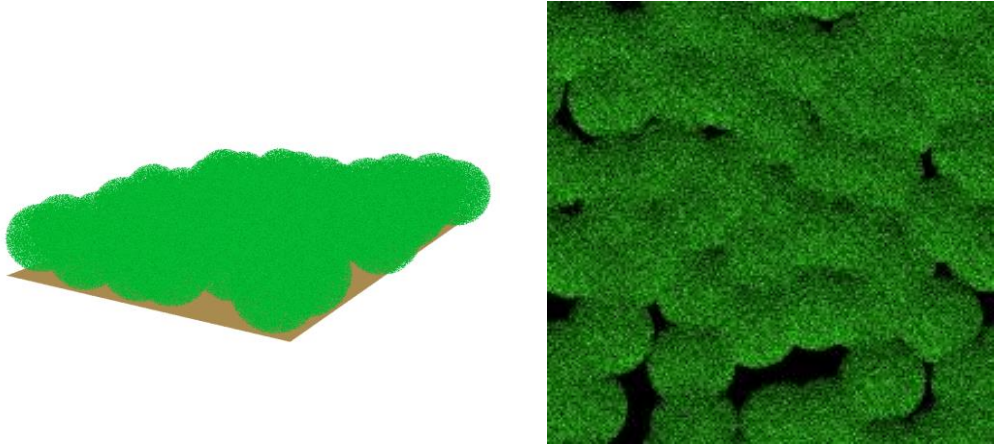

Fig. 6 A 10 m×10 m heterogeneously vegetated scene simulated by DART (a) and its at nadir real color image with a relative azimuth angle of 0°(b). The real color image has a resolution of 0.05 m and is a composite of BRF images in R (650 nm), G (550 nm) and B (450 nm).

Fig. 7 and Tab. 5 show that *iDASF* outperforms *sDASF* consistently in the heterogeneously vegetated scene. The distribution of DASF estimated from *iDASF* approaches both $DASF_0$ and the corresponding non-absorbing canopy BRF. Compared with the results of the 1-D vegetation layer from PROSAIL-D, the advantage of *iDASF* over *sDASF* is reduced but still significant, with the rRMSE reduced by 34%, 30% and 37% for LAI, LIDF and VZA, respectively (Tab. 5). The reduction can be explained by the fact that the *DC* model was built based on the 1-D PROSAIL-D model, without considering the 3-D vegetation structure. Nevertheless, the robustness of the *DC* model still proves to be satisfactory, as the $DC_0$ of the 3-D vegetated scene and the modeled *DC* shows quite good agreement over different canopy structure configurations (Fig. 8).

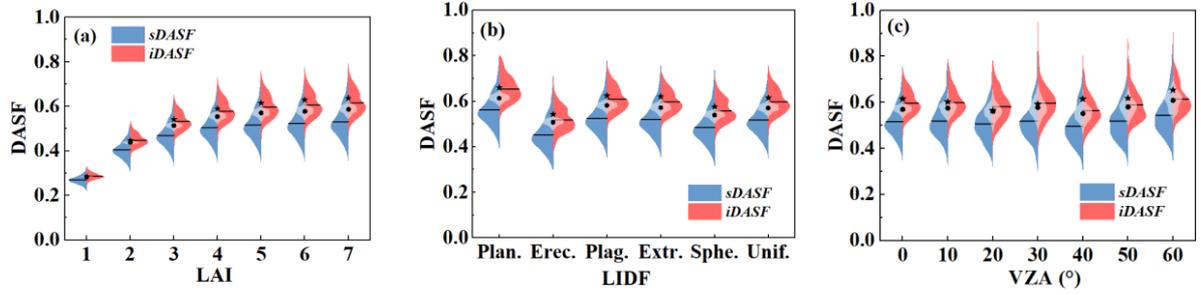

Fig. 7 Split violin distribution of DASF retrieved from the *sDASF* and *iDASF* algorithms. Results were simulated by DART for the scene in Fig. 4, along various (a) LAIs, (b) LIDFs and (c) VZAs. The horizontal lines of the left and right violins denote the mean values. The transparent violin distribution between two split violins represents *DASF$_0$*. The black dots represent mean values of *DASF$_0$*. The black stars denote non-absorbing canopy BRF.

Tab. 5 rRMSE of DASF retrieved from the *sDASF* and *iDASF* algorithms.

| rRMSE (%) | LAI | | | | | | |
|---|---|---|---|---|---|---|---|
| | 1 | 2 | 3 | 4 | 5 | 6 | 7 |
| *sDASF* | 7.76 | 10.56 | 12.22 | 13.12 | 13.57 | 13.74 | 13.87 |
| *iDASF* | 4.47 | 6.44 | 7.97 | 8.92 | 9.51 | 9.72 | 9.92 |
| | LIDF | | | | | | |
| | Plan. | Erec. | Plag. | Extr. | Sphe. | Unif. | |
| *sDASF* | 12.45 | 15.94 | 13.95 | 13.17 | 14.81 | 13.57 | |
| *iDASF* | 10.61 | 9.76 | 9.69 | 9.14 | 9.42 | 9.51 | |
| | VZA (°) | | | | | | |
| | 0 | 10 | 20 | 30 | 40 | 50 | 60 |
| *sDASF* | 13.57 | 13.95 | 14.20 | 15.03 | 14.82 | 15.45 | 15.36 |
| *iDASF* | 9.51 | 9.37 | 9.32 | 9.74 | 8.86 | 8.76 | 8.38 |

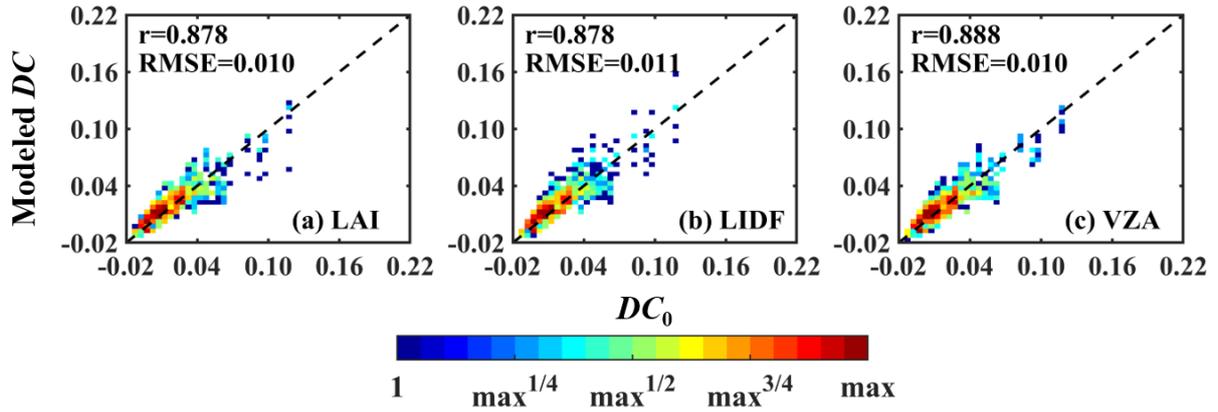

Fig. 8 Scatter plots of $DC_0$ and modeled $DC$ for variation of LAI (a), LIDF (b) and VZA (c) over 3-D heterogeneous vegetation. Each panel illustrates results from all values of the corresponding structural parameter listed in Tab. 5. The values in each colored bin represent the number of scatters on its logarithm scale, so that the bins with warmer colors indicate much more numbers of scatters than the bin with cooler colors.

## 4. Validation of *iDASF* using canopy-level measurements

To validate the *iDASF* algorithm with real canopy-level measurements, we used multi-angular spectral libraries for common European tree species released by Hovi et al. (2021). This dataset consists of simultaneous measurements of canopy BRF and leaf albedo, which makes it possible to obtain $DASF_0$. In this section, we briefly introduces the data acquisition and processing method, and then gives the validation results and the corresponding discussion of *iDASF* using this dataset.

**4.1 Multi-angular reflectance dataset for pine and oak canopies**

From this dataset, Scots pine (*Pinus sylvestris* L.) and Sessile oak (*Quercus petraea* (Matt.) Liebl.) were selected from the spectral library to represent needle leaf and broadleaf species, respectively. For each species, six structurally different canopies were measured indoors to obtain, among others, directional-hemispherical reflectance and transmittance factor (DHRF and DHTF) of needles or leaves and canopy directional scattering coefficient (DSC) (Fig. 9 (a)). The sum of the DHRF and DHTF gave $\omega_\lambda$ for each leaf sample, and their average over two sides and three samples were considered the average canopy $\omega_\lambda$ (Hovi et al. 2020). Multi-angular DSC was obtained using an indoor goniometer, whose viewing angle notation is shown in Fig. 9 (b); observations from a total of 54 viewing directions (46 oblique plus 8 nadir) were taken for each canopy (Hovi et al. 2021). Canopy BRF was obtained by multiplying DSC by $\pi$ (Hovi et al. 2021). More details about the experimental condition and data processing can be found in Hovi et al. (2021) and Forsstrom et al. (2021).

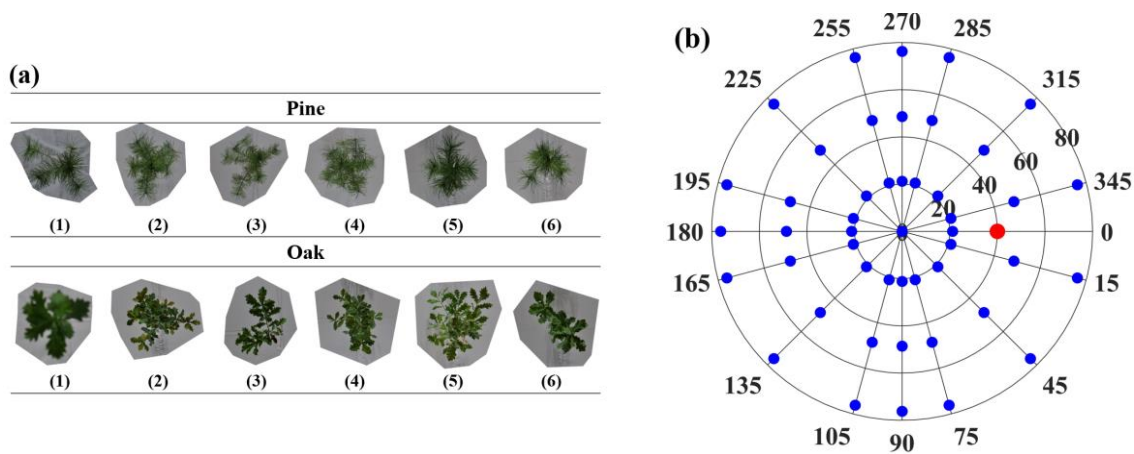

Fig. 9 (a) Photos of canopies recorded in Hovi et al. (2021) for which the multi-angle scattering spectra were measured. (b) Viewing angle notation for the data collection. Blue and red circles denote the viewing angles and the illumination position, respectively.

## 4.2 Improvement of DASF estimation using *iDASF*

We calculated the absolute error (AE) between the estimated and measured DASF ($DASF_0$) to evaluate the accuracy of the *sDASF* and *iDASF* algorithms. The probability distribution of AE calculated from 324 observations (54 multiplied by 6 canopies) per species (Fig. 10) shows the improvement in accuracy achieved by *iDASF*. For pine canopies (Fig 10 (a)), *sDASF* shows a largely biased estimation with a mean absolute error (MAE) of 0.022, whereas *iDASF* produces DASF with an error distribution much closer to zero and a MAE of 0.007, 68% lower compared to *sDASF*. The same result but with less improvement was observed for oak canopies. The MAE is reduced from 0.005 to 0.004 (by 20%).

We then calculated the difference between AE obtained from *iDASF* and *sDASF* and showed its averaged variation with viewing geometries over six canopies (Fig 11). For pine canopies, *iDASF* gives significant and stable improvements, with AE reduced by more than 0.01 for all the viewing angles. For oak canopies, *iDASF* slightly improved the estimation accuracy for most of the viewing angles, except for angles at 76.2° zenith. In such cases, *iDASF* produced either similar (yellow grids) or slightly worse (orange to red grids) performance compared to *sDASF*. Such distribution pattern is closely related to that of $DASF_0$, which is shown in Fig. 12. Except for near-hot-spot regions, the DASF of pines is uniformly distributed along the viewing directions (Fig 12 (a)), whereas, for oaks, 76.2° VZA generally yields lower values of DASF (Fig 12(b)). This can be explained by the fact that most of the leaves within the studied oak canopies are distributed horizontally(Hovi et al. 2021). As a result, there is a decrease in the proportion of leaves that forms the canopy boundary in total leaf area when VZA is large, which causes the large decline in DASF. This effect, however, does not appear in pine canopies.

A factor that would essentially affect the improvement of *iDASF* on oaks could be the concentrations of $C_{ab}$ and LMA. The fact is that the LMA of a needle is usually higher than that of a broadleaf (Knyazikhin et al. 2013), whereas $C_{ab}$ of pines is usually lower than that of healthy oaks (without leaf senescence and abscission)(Eitel et al. 2011; Wong et al. 2019). This means that for pine species, the *sDASF* assumption that variation of leaf scattering albedo with dry matter can be neglected is more difficult to be satisfied. In other words, LMA has greater impact on variation of leaf albedo for pines, and the ratio of real LMA and reference LMA (0.005 $g/cm^2$) may be much larger than the ratio of real $C_{ab}$ and reference $C_{ab}$ (16 $\mu g/cm^2$). As was mentioned in Section 2.3, *sDASF* seriously underestimates DASF in such cases (Fig. 10 (a)). In comparison, a biased estimation is not that obvious for oak canopies, only leading to uncertainty in the estimation of DASF for oaks using *sDASF*. By adopting *iDASF*, the underestimation for pine canopies was eliminated and more accurate results for oak canopies were produced.

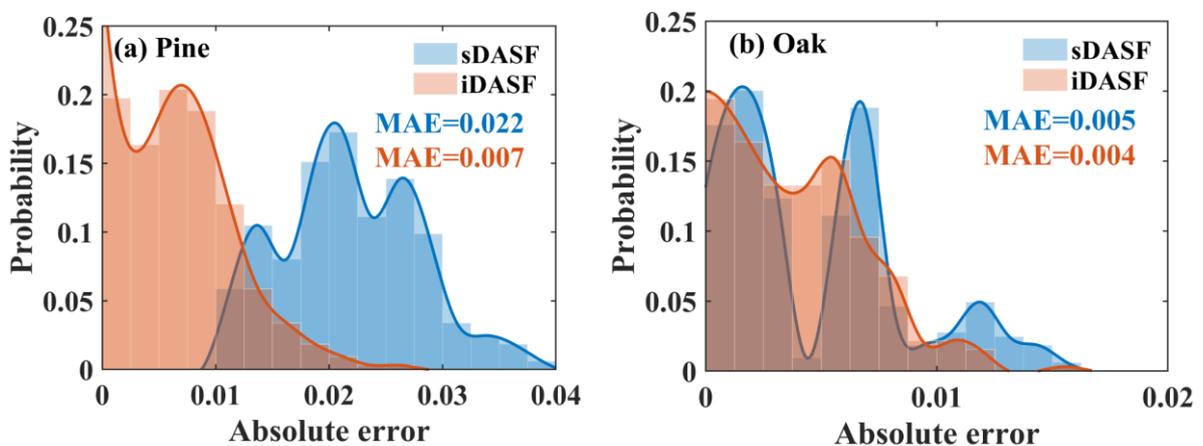

Fig. 10 Histogram and corresponding probability distribution of the absolute error of estimated DASF

from the *sDASF* and *iDASF* algorithms for all the 324 observations of pine (a) and oak (b) canopies.

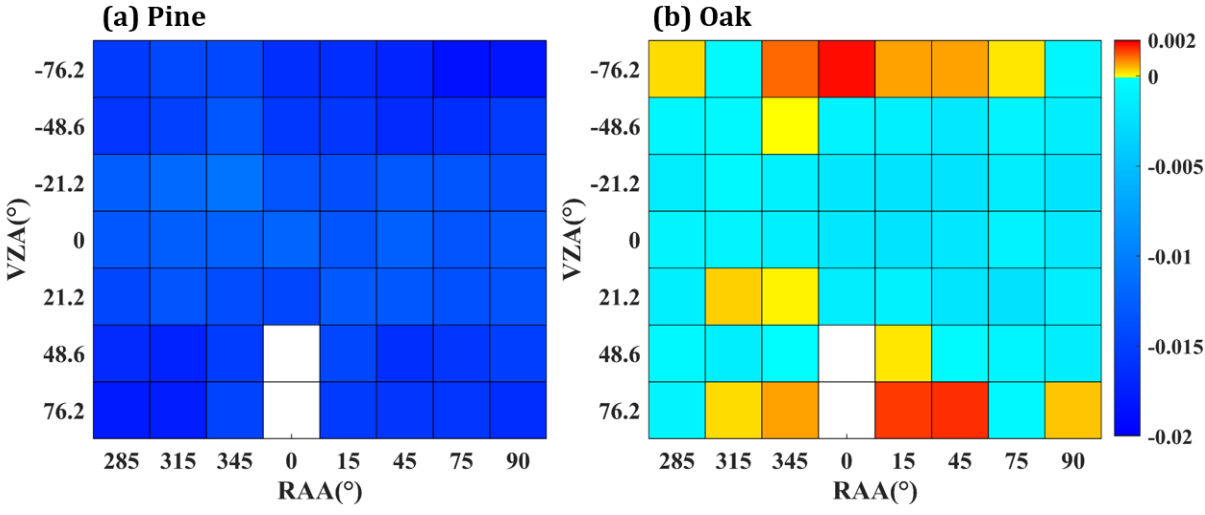

Fig. 11 Averaged difference between the absolute error obtained from *iDASF* and from *sDASF* over six pine (a) and oak (b) canopies along view zenith angles and relative azimuth angles shown in Fig. 9(b). For VZA, positive degrees represent backscattering directions, whereas negative degrees represent forward scattering directions. The two blank grids denote the hot-spot region where measurements were not taken. Grids in blue indicate better performance of *iDASF* and those in red indicate better performance of *sDASF*. Yellow grids suggest a similar performance of *iDASF* and *sDASF*, with a negligible difference of AE.

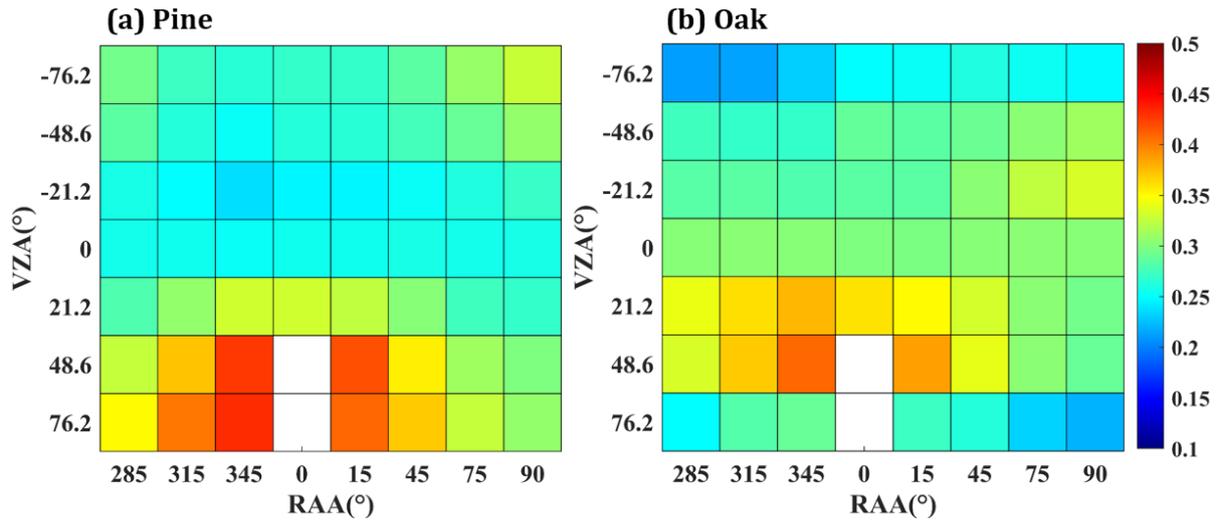

Fig. 12 Averaged *DASF*$_0$ over six pine (a) and oak (b) canopies along view zenith angles and relative azimuth angles shown in Fig. 9(b).

## 5. Conclusion

Accurate estimation of DASF is an essential task for many cases of quantitative vegetation monitoring. Because of the unsatisfied assumption that variation of leaf scattering albedo with dry matter is neglected, which is the basic premise in the widely used standard approach for DASF retrieval (*sDASF*), this study proposed an improved DASF estimation algorithm (*iDASF*), simple and easy to use, by incorporating variation of leaf dry matter. *iDASF* outperformed *sDASF* on both a 1-D homogeneous vegetation layer and a 3-D heterogeneous vegetated scene, and was not sensitive to various canopy structural characteristics. The advantage of *iDASF* was further confirmed by a recently released canopy multi-angular spectral library. More accurate DASF estimates can lead to more effective decoupling of canopy structure and leaf optics. More analyses of the relationship between leaf biochemistry (e.g., leaf nitrogen content) and canopy scattering signal corrected for canopy structure (e.g., canopy scattering coefficient) should be conducted in the future to verify this. The *iDASF* algorithm could enhance the application potential of DASF and provide a more accurate and comprehensive understanding of changes in vegetation functional traits, dynamics, and other key parameters of terrestrial ecosystem.

## Credit author statement

Yi Lin: Conceptualization, Review, Funding acquisition, Project administration, Supervision. Siyuan Liu: Methodology, Software, Data analysis, Evaluation, Writing, Visualization. Lei Yan: Investigation, Review, Funding acquisition, Project administration, Supervision. Kai Yan and Yelu Zeng: Review, Formal analyses, Editing. Bin Yang: Conceptualization, Investigation, Methodology, Writing, Review, Editing, Funding acquisition, Project administration.

## Declaration of Competing Interest

The authors declare that they have no known competing financial interests or personal relationships that could have appeared to influence the work reported in this paper.

## Acknowledgements

This work was supported by the National Natural Science Foundation of China (Grant No. 41801227, 32171782 and 31870531) and National Key R&D Program of China (Grant No. 2017YFB0503004). We are grateful to Professor Yuri Knyazikhin from Boston

University for the very constructive discussions and suggestions on this work. We also thank Yingjie Wang and Professor Jean-Philippe Gastellu-Etchegorry from CESBIO, CNES-CNRS-IRD-UPS, University of Toulouse, for their provision of the *HET 20* vegetated scene used in RAMI-V and all their help regarding the DART software.

# Appendix

## A.1 Spectrally invariant relationships for leaf albedo at weakly absorbing wavelengths

The transformed leaf albedo is defined as the conditional probability of photons scattered by the leaf given that they interact with leaf internal constituents (Lewis and Disney 2007). In the weakly absorbing 710 to 790nm spectral interval, the transformed leaf albedo can be expressed via a fixed spectrally varying reference leaf albedo $\varpi_r$, as (Knyazikhin et al. 2013; Latorre Carmona et al. 2010; Schull et al. 2011)

$$\varpi_\lambda = \frac{r(Cab, LMA)}{1 - p(Cab, LMA)\varpi_r} \varpi_r \ . \tag{A1}$$

Here $\varpi_r$ is the reference albedo generated using PROSPECT model with leaf chlorophyll concentration ($C_{ab}$) 16 $\mu g/cm^2$, equivalent water thickness 0.005 $cm^{-1}$ and leaf dry matter per unit area (LMA) 0.002 $g/cm^2$; $p(Cab, LMA)$ denotes the within-leaf recollision probability, $r(Cab, LMA) + p(Cab, LMA)=1 + \varepsilon$, where $\varepsilon$ characterizing the deviation of the left hand sum from unity, and $\lambda$ stands for wavelength. The within-leaf recollision probability depends on concentrations of $C_{ab}$ and LMA. Absorption by other biochemical constituencies is negligibly small in this spectral interval. The standard technique provides unbiased estimates of DASF if and only if $\varepsilon$=0.

Analysis of Equation (A1) based on synthetic dataset (Section 2.3) suggests that $p(C_{ab}, LMA)$ is linearly related to the within-leaf recollision probability $p_0(C_{ab}) = p(C_{ab}, 0.002 g/cm^2)$ with LMA=$0.002 g/cm^2$, i.e.,

$$p(Cab, LMA) = k(LMA) \cdot p_0(C_{ab}) + b(LMA) \quad . \tag{A2}$$

Here $k$ and $b$ are linear functions of LMA, while $p_0$ varies inversely with $C_{ab}$, namely (Fig. A1)

$$k(LMA) = 9.18 \cdot LMA + 0.98 \tag{A3}$$

$$b(LMA) = -9.16 \cdot LMA + 0.02 \tag{A4}$$

$$p_0(C_{ab}) = 1.04 - \frac{15.54}{C_{ab}} \tag{A5}$$

Fig. A1 shows that $k(LMA)$ and $b(LMA)$ exhibit relatively weak variations with $LMA$: the former varies between 0.99 and 1.13 with a mean (standard deviation) of 1.07 (0.05) while the latter takes values in the range -0.13 to 0.01 with a mean (standard deviation) of -0.07 (0.05). The within-leaf recollision probability can thus be approximated using their mean values as

$$p(Cab, LMA) = 1.07 \cdot p_0(C_{ab}) - 0.07 = 1.04 - \frac{16.63}{C_{ab}}. \tag{A6}$$

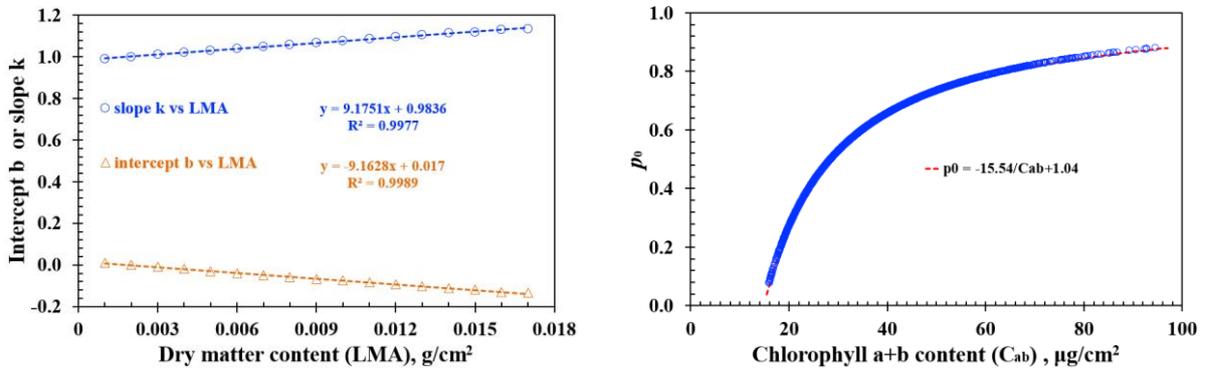

**Figure A1.** (a) Relationship between $k(LMA)$, $b(LMA)$ and dry matter content. (b) Relationship between $p_0(C_{ab})$ and $C_{ab}$.

Analogously, $r(C_{ab}, LMA)$ is also linearly related to $p_0(C_{ab})$. The relationship between $r$ and $C_{ab}$ can be approximated as

$$r(Cab, LMA) = -0.97 \cdot p_0(C_{ab}) + 0.99 = \frac{15.07}{C_{ab}} - 0.02. \tag{A7}$$

The deviation $\varepsilon$ therefore can be estimated as $\varepsilon \sim 0.1 p_0(C_{ab}) - 0.08$, which varies between -0.07 and +0.01.

If the concentration of leaf chlorophyll is expressed relative to the chlorophyll content in the reference leaf albedo, i.e., $C_{ab} = 16t\ \mu g/cm^2$, the within-leaf recollision probability rearranges to $p(C_{ab}, LMA) \approx (t-1)/t$. If one increases the reference concentration of dry matter $t$ times, the deviation $\varepsilon(16t, 0.002t)$ will take the value 0 and, $r(C_{ab}, LMA)$ consequently becomes $1/t$ (Yang et al. 2016). In all other cases, $\varepsilon \neq 0$ and, as estimated above, its variation does not exceed 7%. The standard technique assumes negligible impact of variation in amount of leaf dry matter on estimation of DASF.

Incorporating transformed leaf albedo with variation of leaf dry matter can help to tackle this problem, which is shown in Appendix A.2 and A.3.

## A.2. Estimation of transformed leaf albedo of green leaves without neglecting LMA

*In silico* analysis of the transformed leaf albedo $\varpi_\lambda$ based on the leaf scattering PROSPECT model suggests that $\varpi_\lambda$ can be represented as (Lewis and Disney 2007):

$$\varpi_\lambda = \frac{(1 - p_{leaf})W_{leaf}}{1 - p_{leaf}W_{leaf}}, \tag{A8}$$

where $W_{leaf}$ denotes a fundamental scattering term defined as $\exp(-\sum_{i=1}^{i=m} C_i k_i(\lambda))$ and $p_{leaf}$ is the recollision probability between the within-leaf fundamental terms. Here $C_i$ stands for the concentration of $i$th biochemical constituent, $k_i(\lambda)$ denotes modified a specific absorption coefficient, and $m$ is the number of biochemical constituents present in the leaf. $p_{leaf}$ is wavelength independent, whereas $W_{leaf}$ and $\varpi_\lambda$ are spectrally varying quantities.

In the 710 to 790 nm spectral region, leaf scattering property is mainly determined by the absorption of chlorophyll and leaf dry matter. The former varies with wavelength, whereas the latter is flat. Absorption by other biochemical constituencies are negligibly small (Feret et al. 2008). Now we introduce the reference transformed leaf albedo $\varpi_r$. Let $C_{ab}$ and $C_m$ be the concentrations of chlorophyll and dry matter of the reference leaf, respectively; $k_c(\lambda)$ and wavelength independent $k_m$ be their modified specific absorption coefficients. The fundamental scattering term of the reference leaf $W_{leaf,r}$ can be calculated as $\exp(-C_{ab}k_c(\lambda) - C_m k_m)$ and the reference transformed leaf albedo is $\varpi_r = \frac{(1-p_{leaf})W_{leaf,r}}{1-p_{leaf}W_{leaf,r}}$. $W_{leaf,r}$ and $\varpi_r$ are related as

$$W_{leaf,r} = \frac{\varpi_r}{1 - p_{leaf} + p_{leaf}\varpi_r}. \tag{A9}$$

Consider a green leaf whose concentrations of chlorophyll and dry matter are $t_c C_{ab}$ and $t_m C_m$, the corresponding fundamental scattering term $W_{leaf}$ is,

$$W_{leaf} = \exp(-t_c C_{ab} k_c(\lambda) - t_m C_m k_m) = A W_{leaf,r}^{t_c}, \tag{A10}$$

where spectrally invariant coefficient has the form $A = \exp(-(t_m - t_c)C_m k_m)$. The transformed leaf albedo of the green leaf $\omega_\lambda$ thus rearranges to the following form:

$$\varpi_\lambda = \frac{(1-p_{leaf})AW_{leaf,r}^{t_c}}{1 - p_{leaf}AW_{leaf,r}^{t_c}}, \tag{A11}$$

In the 710-790 nm spectral region, the term $W_{leaf,r}^{t_c}$ can be accurately approximated as (Yang et al. 2016)

$$W_{leaf,r}^{t_c} = \frac{1 - q(t_c)}{1 - q(t_c)W_{leaf,r}} W_{leaf,r}, \tag{A12}$$

where $q(t_c) = (t-1)/t$. Substitution of Equation (A12) and Equation (A9) into Equation (A11) results in

$$\varpi_\lambda = \frac{A[1 - q(t_c)]}{1 - B\varpi_r} \varpi_r, \tag{A13}$$

where,

$$B = \frac{q(t_c) - p_{leaf} + p_{leaf}A[1 - q(t_c)]}{1 - p_{leaf}}. \tag{A14}$$

The variable $B$ is independent of wavelength. If variation in $\varpi_\lambda$ with LMA is neglected, then $A = 1$ and $B = q(t_c)$, and Equation (A13) becomes

$$\varpi_\lambda = \frac{1 - q(t_c)}{1 - q(t_c)\varpi_r} \varpi_r, \tag{A15}$$

## A.3. Biased estimation of DASF using the reference leaf albedo

According to the spectral invariants theory, if the contribution of canopy understory on canopy BRF is negligible, the canopy BRF can be simply parameterized as a function of leaf albedo $\omega_\lambda$, canopy interceptance $i_0$, recollision probability $p$, and directional escape probability $\rho(\Omega)$, as (Huang et al. 2007; Knyazikhin et al. 2011),

$$BRF_\lambda(\Omega) = \frac{\rho(\Omega)i_0}{1 - p\omega_\lambda} \omega_\lambda. \tag{A16}$$

The leaf albedo $\omega_\lambda$ results from two components, i.e., photons that interact with the leaf surface and the interior:

$$\omega_\lambda = s_L + (1 - s_L)\varpi_\lambda. \tag{A17}$$

where $s_L$ is the fraction of photons reflected at the leaf surface. The variable is wavelength insensitive and varies with the leaf surface properties. It is usually small (on the order of $10^{-2}$) (Lewis and Disney 2007). Here we neglect $s_L$ and approximate the leaf albedo with $\varpi_\lambda$, i.e., $\omega_\lambda \approx \varpi_\lambda$.

If variation in the within-leaf recollision probability with dry matter content is ignored, the BRF can be calculated by substitution of Equation (A15) to Equation (A16),

$$BRF_\lambda(\Omega) = \frac{\rho(\Omega)i_0[1-q(t)]}{1-[q(t)+p(1-q(t))]\varpi_r}\varpi_r. \tag{A18}$$

By plotting $BRF_\lambda(\Omega)/\varpi_r$ versus $BRF_\lambda(\Omega)$, a linear relationship is achieved, where the intercept is $\rho(\Omega)i_0[1-q(t)]$, and the slope is $q(t)+p(1-q(t))$. The ratio between the intercept and (1-slope) becomes independent of $q(t)$ and $\varpi_r$, and provides DASF as,

$$DASF = \frac{\rho(\Omega)i_0}{1-p}. \tag{A19}$$

Incorporating variation of dry matter content, a true BRF can be calculated by substituting Equation (A13) into Equation (A16),

$$BRF_\lambda(\Omega) = \frac{A\rho(\Omega)i_0[1-q(t_c)]}{1-[B+pA(1-q(t_c))]\varpi_r}\varpi_r. \tag{A20}$$

If we follow the standard technique and plot the $BRF_\lambda(\Omega)/\varpi_r$ versus $BRF_\lambda(\Omega)$, a linear relationship can still be achieved, where the intercept and slope are $A\rho(\Omega)i_0[1-q(t_c)]$ and $B+pA(1-q(t_c))$, respectively. The ratio between the intercept and (1-slope) gives a new DASF, namely, $DASF'$,

$$DASF' = \frac{D\rho(\Omega)i_0}{1-B-Dp}. \tag{A21}$$

where,

$$D = A[1-q(t_c)]. \tag{A22}$$

By substituting Equation (A14) to Equation (A21), the latter can be reorganized as,

$$DASF' = \frac{D\rho(\Omega)i_0}{D(1-p)+DC} = \frac{\rho(\Omega)i_0}{1-p+C}. \tag{A23}$$

where,

$$C = \frac{1-A}{A(1-p_{leaf})}. \tag{A24}$$

Moreover, if variation of LMA is neglected, then $C = 0$, and $DASF'$ is identical to true DASF. Otherwise, comparison between Equation (A19) and Equation (A23) suggests that the standard technique of DASF retrieving gives $DASF'$, but not the true DASF.

## References

Adams, J., Lewis, P., & Disney, M. (2018). Decoupling Canopy Structure and Leaf Biochemistry: Testing the Utility of Directional Area Scattering Factor (DASF). *Remote Sens., 10*

Brown, L.A., Fernandes, R., Djamai, N., Meier, C., Gobron, N., Morris, H., Canisius, F., Bai, G., Lerebourg, C., Lanconelli, C., Clerici, M., & Dash, J. (2021). Validation of baseline and modified Sentinel-2 Level 2 Prototype Processor leaf area index retrievals over the United States. *ISPRS-J. Photogramm. Remote Sens., 175*, 71-87

Clevers, J.G.P.W., & Kooistra, L. (2012). Using Hyperspectral Remote Sensing Data for Retrieving Canopy Chlorophyll and Nitrogen Content. *IEEE J. Sel. Top. Appl. Earth Observ. Remote Sens., 5*, 574-583

Eitel, J.U.H., Vierling, L.A., Long, D.S., Litvak, M., & Eitel, K.C.B. (2011). Simple assessment of needleleaf and broadleaf chlorophyll content using a flatbed color scanner. *Can. J. For. Res., 41*, 1445-1451

Feret, J.B., Berger, K., de Boissieu, F., & Malenovsky, Z. (2021). PROSPECT-PRO for estimating content of nitrogen-containing leaf proteins and other carbon-based constituents. *Remote Sens. Environ., 252*

Feret, J.B., Francois, C., Asner, G.P., Gitelson, A.A., Martin, R.E., Bidel, L.P.R., Ustin, S.L., le Maire, G., & Jacquemoud, S. (2008). PROSPECT-4 and 5: Advances in the leaf optical properties model separating photosynthetic pigments. *Remote Sens. Environ., 112*, 3030-3043

Feret, J.B., Francois, C., Gitelson, A., Asner, G.P., Barry, K.M., Panigada, C., Richardson, A.D., & Jacquemoud, S. (2011). Optimizing spectral indices and chemometric analysis of leaf chemical properties using radiative transfer modeling. *Remote Sens. Environ., 115*, 2742-2750

Féret, J.B., Gitelson, A.A., Noble, S.D., & Jacquemoud, S. (2017). PROSPECT-D: Towards modeling leaf optical properties through a complete lifecycle. *Remote Sens. Environ., 193*, 204-215